\RequirePackage{fix-cm}
\documentclass[twocolumn]{svjour3}          
\smartqed  
\usepackage{graphicx}
\begin{document}

\title{Estimation of intraband and interband relative coupling constants from temperature dependences of the order parameter for two-gap superconductors}

\titlerunning{Intraband to interband coupling rate for two-gap superconductors}        

\author{S.A. Kuzmichev
\and T.E. Kuzmicheva
\and S.N. Tchesnokov
\and V.M. Pudalov
\and A.N. Vasiliev}


\institute{S.A. Kuzmichev \and S.N. Tchesnokov \and A.N. Vasiliev \at
              Physics Faculty, M.V. Lomonosov Moscow State University, Moscow \\
              119991, Russia
              \email{kuzmichev@mig.phys.msu.ru}           
           \and
              T.E. Kuzmicheva \and V.M. Pudalov \at
              P.N. Lebedev Physical Institute of RAS, Moscow \\
              119991, Russia
              \and
              A.N. Vasiliev \at
              Theoretical Physics and Applied Mathematics Department, Ural Federal University, 620002 Ekaterinburg, Russia \\
              National University of Science and Technology "MISiS", Moscow 119049, Russia
}

\date{Received: date / Accepted: date}

\maketitle

\begin{abstract}
We present temperature dependences of the large and the small superconducting gaps measured directly by SnS-Andreev spectroscopy in various Fe-based superconductors and MgB$_2$. The experimental $\Delta_{L,S}(T)$ are well-fitted with a two-gap model based on Moskalenko and Suhl system of equations (supplemented with a BCS-integral renormalization). From the the fitting procedure, we estimate the key attribute of superconducting state \textemdash relative electron-boson coupling constants and eigen BCS-ratios for both condensates. Our results evidence for a driving role of a strong intraband coupling in the bands with the large gap, whereas interband coupling is rather weak for all the superconductors under study.
\end{abstract}

\section{Introduction}
In spite of the intensive theoretical and experimental studies  stimulated by the recent discovery of Fe-based superconductors in 2008 \cite{Kamihara}, many aspects of the multiband superconducting state are not fully understood yet. Particularly, one of the candidates for pairing in Fe-HTS is the spin-fluctuation-based pairing mechanism that leads to the $s^{\pm}$ symmetry of the order parameter \cite{Mazin}. A different mechanism, where pairing is mediated by orbital fluctuations \cite{Onari}, leads to the $s^{++}$ state.

Among the iron-based superconductors, the 1111 family have the simplest band structure \cite{Nekrasov,Kuroki}. The bands crossing the Fermi level form quasi-two-dimensional sheets at the Fermi surface. According to recent ARPES studies, they can be considered as two effective electron and hole bands where the large (L) and the small (S) superconducting gaps are developed at $T < T_C$ \cite{Charnukha}. However, despite the apparent simplicity, the majority of the experimental techniques face with serious troubles probing the values of the order parameter in 1111. The absence of single crystals of sufficient dimensions, and a charged surface of cryogenic clefts \cite{Yin,Zhigadlo2} strongly distort the experimental data. As a result, the values of superconducting gaps and corresponding BCS-ratios are very contradictive: for oxypnictides, $2\Delta_L/k_BT_C$ vary more than by a factor of 6 (for a review, see \cite{InosovRev,UFN} and Refs. therein), from BCS-limit 3.5 up to 22 \cite{GonnRu}. Temperature dependences of the gaps are also ambiguous (the data are reviewed in \cite{UFN,Sm}, see also Refs. therein). For example, in \cite{DagheroRev} a BCS-like temperature dependence was found for both gaps in SmO(F)FeAs, whereas in LaO(F)FeAs the large gap turned to zero at $T \approx 2/3 T_C$ while the small gap was closed quite linearly and survived till the $T_C$. The latter was recognized as an artifact \cite{DagheroRev} and was reproduced nowhere in literature.

Fortunately, the 1111 oxypnictide superconductors are good candidates for probing by multiple Andreev reflections effect \cite{Andreev} spectroscopy. Here we present directly determined  temperature dependences of the large and the small gaps in various 1111 compounds with the wide range of critical temperatures $21.5 {\rm K} \leq T_C \leq 50 {\rm K}$. The measured $\Delta_{L,S}(T)$ dependences were fitted using a two-band BCS-like model with a renormalized BCS (RBCS) integral based on Moskalenko and Suhl system of gap equations \cite{Mosk,Suhl,MgB2fit}. The relative values of electron-boson coupling constants $\lambda_{i,j}/\lambda_{LL}$ ($i,j = L,S$) are estimated. The data for 1111 oxypnictides are compared with $\Delta(T)$ and $\lambda_{i,j}$ for LiFeAs and Mg$_{1-x}$Al$_x$B$_2$.

\section{Experimental details}

We used LiFeAs single crystals with critical temperature $T_C^{bulk} \approx 16 \textendash 17$\,K (the synthesis is detailed in \cite{Morozov}), and the following polycrystalline samples: fluorine-doped LaO$_{1-x}$F$_x$FeAs with $T_C^{bulk} \approx 22 \textendash 28$\,K \cite{Kondrat,LOFA}, Sm$_{1-x}$Th$_x$OFeAs with a wide range of thorium doping and critical temperatures $T_C^{bulk} \approx 25 \textendash 55$\,K \cite{Zhigadlo2,Zhigadlo1}, optimal oxygen-deficient GdO$_{0.88}$FeAs with $T_C^{bulk} \approx 50$\,K \cite{Khlybov}, and Mg$_{1-x}$Al$_x$B$_2$ with $T_C^{bulk} \approx 21.5 \textendash 40$\,K \cite{Kr,Bulychev}. The common feature of the samples under study is a layered crystal structure.

In our studies, superconducting gaps and $\Delta(T)$ temperature dependences were probed by Andreev spectroscopy of superconductor - normal metal - superconductor (SnS) contacts \cite{Andreev}, and intrinsic multiple Andreev reflections effect (IMARE) spectroscopy \cite{Pon_IMARE,EPL}. SnS-contacts were formed by precise cleaving of the superconducting samples at $T=4.2$\,K using a break-junction technique \cite{Moreland}. Under such cleavage, a layered sample exfoliates along the $ab$-planes; as a result, it contains two cryogenic clefts separated by a weak-link. For Fe-based superconductors of the 1111, and 111 families \cite{Sm,EPL,GdJETPL,GdJPCS,Li12,Li13}, the current-voltage characteristic (CVC) and the dynamic conductance correspond to a ballistic \cite{Sharvin} high-transparent SnS-Andreev contact \cite{OTBK,Arnold,Kummel}. The break-junction technique enables the formation of clean cryogenic clefts, true 4-probes connection, thus preventing overheating of the contact area, and easy mechanical readjustment. It is well-applicable for both single crystals and polycrystalline samples of layered material \cite{EPL}. Our dI(V)/dV measurements were performed directly by a modulation technique. We used a current source with ac frequency less than 1\,kHz. The results obtained with this setup are insensitive to the potential presence of  parallel ohmic conduction paths; the latter, if present, only shift dynamic conductance curves along the vertical axis.

The multiple Andreev reflections effect occurring in ballistic SnS interface manifests itself as a pronounced excess current at low bias voltages (so-called foot area), and a series of dynamic conductance dips at certain positions $V_n = 2\Delta/en$, where $n$ is a natural subharmonic order; these features are called subharmonic gap structure (SGS). The gap value may be directly determined from the SGS positions over all the temperature range from 0 to $T_C$ with no dI(V)/dV fitting \cite{Kummel}. In case of two-gap superconductor, two such SGS's should be observed. Along with multiple Andreev reflections effect in single SnS-contacts, in layered samples we observe an intrinsic multiple Andreev reflections effect (IMARE) in the stack structures of S-n-S-n-\dots-S-type \cite{UFN,Sm,EPL,Li12,Li13}. The IMARE is similar to intrinsic Josephson effect \cite{Nakamura}, and was first observed in Bi cuprate superconductors \cite{IJE}. Since the array of Andreev contacts represents a sequence of $m$ identical SnS junctions, the SGS appears at $V_n = m \cdot2\Delta/en$ bias voltages. We have shown that with $m$ increasing, the contribution of parasitic effects decreases, facilitating observation of bulk superconducting gaps \cite{EPL}.

\section{Results and discussion}

Figure 1 shows normalized dynamic conductance curves for Andreev array (2 junctions in the stack) for LaO$_{1-x}$F$_x$FeAs sample. Note that the absolute dynamic conductance decreases with temperature, while the dI(V)/dV spectra in Fig. 1 are shown in arbitrary units, and offset vertically for clarity. The inset of Fig. 2 shows current-voltage characteristic (CVC) for this contact at $T = 4.2$\,K with a pronounced excess current at low biases (foot). The contact has a local critical temperature $T_C^{local} \approx 22$\,K corresponding to the contact area transition to the normal state; above this temperature the CVC becomes ohmic-like and free of an excess conductance. Taking the known product of quasiparticle mean free path and normal-state resistivity $\rho l \approx 10^{-10}$\,${\rm \Omega cm^2}$ for 1111 compounds \cite{LOFA,Shan,Zhu}, we get $l \approx 50$\,nm. Then, using Sharvin formula \cite{Sharvin}, and the resistance of this contact $R \approx 28$\,${\rm \Omega}$ (here and below all calculations refer to a single junction in the array), we estimate the contact diameter $a = \sqrt{4/3\pi \times \rho l/R} \approx 12$\,nm $ \ll l$. The average crystallite dimensions are $60 \textendash 70 {\rm \mu m}$ in Sm-1111 samples under study, and $10 \textendash 40 {\rm \mu m}$ in La-1111 samples. These values are much larger than the estimated contact diameter $a$, thus providing local study of crystallites. Since $a \ll l$, and the resistance of this contact decreases with temperature increasing, we conclude that our measurements are in the clean ballistic SnS-Andreev mode \cite{Klapwijk}. The same could be concluded for Mg$_{1-x}$Al$_x$B$_2$ samples we used. The estimated $\rho l$ range is $(2 \textendash 5) \times 10^{-12} {\rm \Omega cm^2}$, and $l \approx 80$\,nm \cite{Eltsev1,Eltsev2}. For our SnS-contacts, $R = 1 \textendash 22 {\rm \Omega}$, which leads to the estimate of the contact diameter $a = 2 \textendash 15$\,nm. Obviously, $a < l$, and less than the typical grain dimensions of about 100\,nm for MgB$_2$ \cite{Bulychev,Kr}. On an appearance of any parallel ohmic contact, the dynamic conductance in a sample circuit increases, which could lead to underestimation of the Sharvin contact resistance. Since $a \sim 1/\sqrt{R}$ in a ballistic contact, true diameter of Andreev contact may be smaller than the above estimate.

\begin{figure}
\includegraphics[width=.49\textwidth]{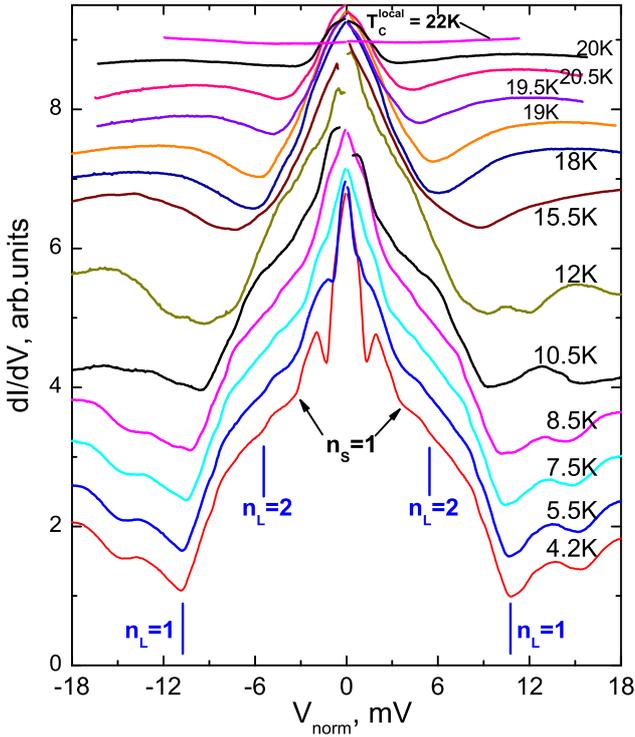}
\caption{Dynamic conductance spectra for SnS Andreev contact in LaO$_{1-x}$F$_x$FeAs with local critical temperature $T_C^{local} \approx 22$\,K. The curves are offset vertically for clarity. At $T=4.2$\,K, $\Delta_L \approx 5.4$\,meV, $\Delta_S \approx 1.4$\,meV, subharmonic gap peculiarities for $\Delta_L$ are marked with vertical dashes and $n_L=1,2$ labels, for $\Delta_S$ \textemdash by arrows and $n_S=1$ label.}
\end{figure}

At $T=4.2$\,K, dynamic conductance peculiarities located at $|V_1| \approx 10.8$\,mV and $|V_2| \approx 5.4$\,mV (marked in Fig. 1 as $n_L=1$ and $n_L=2$, respectively) correspond to the first and the second Andreev minima related with the large gap $\Delta_L = 5.4 \pm 0.5$\,meV. Andreev peculiarities for the small gap $\Delta_S = 1.4 \pm 0.3$\,meV are marked with $n_S=1$ labels and arrows. These values are close to those obtained by Andreev spectroscopy previously \cite{LOFA,FPS11}. Dips at $|V| \sim 1.4$\,mV related to the beginning of the foot area are sharp at $T=4.2$\,K and then becoming smeared with temperature. With temperature increasing, the SGS peculiarities for both gaps move towards zero bias, and at $T_C^{local} \approx 22$\,K the dI(V)/dV curve become linear. Taking the position of SGS peculiarities for the large and the small gap, we directly obtain temperature dependences $\Delta_{L,S}(T)$ presented in Fig. 2 (by solid circles for $\Delta_L(T)$, and by open circles for $\Delta_S$). The temperature behaviour of the large and the small gap for LaO$_{1-x}$F$_x$FeAs looks like that for Sm$_{1-x}$Th$_x$OFeAs shown in Fig. 3, and for other Fe-based superconductors \cite{UFN,Sm,GdJPCS,Li12,Li13,GdJSNM,Ba,KFeSe}.

A two-band BCS-like superconducting system is described by a matrix of four electron-boson coupling constants $\lambda_{ij} = V_{ij} N_j$, $i,j = L,S$ ($L$ index corresponds to the bands with the large gap, $S$ index \textemdash to the bands with the small gap), where $V_{ij}$ is the interaction matrix element between $i^{th}$ and $j^{th}$ bands, $N_j$ is the normal-state density of states at the Fermi level in $j^{th}$ band \cite{Mosk,Suhl}. Varying the relation between intraband and interband coupling, one may model various temperature dependences of the gaps. Obviously, in a case of zero interband interaction ($V_{LS}=V_{SL}=0$), the large and the small gaps follow BCS-type curve and close each at its eigen critical temperature $T_C^{L,S}$ \cite{Nicol,Pickett}. By contrast, in Fe-based superconductors \cite{UFN,Sm,GdJPCS,Li12,Li13,GdJSNM,Ba,KFeSe} as well as in Mg$_{1-x}$Al$_x$B$_2$ \cite{MgB2fit,MgB2JETPL04,MgB2SSC04} within the wide range of critical temperatures we observe the $\Delta_{L,S}(T)$ deviation from the single-gap BCS-like dependences: $\Delta_L(T)$ follows BCS type at the whole, but slightly bends down, whereas $\Delta_S(T)$ at first falls rapidly, then flattens, and finally turns to zero at $T_C^{local}$. As a result, the two gaps close at the common critical temperature $T_C^{local}$.

\begin{figure}
\includegraphics[width=.49\textwidth]{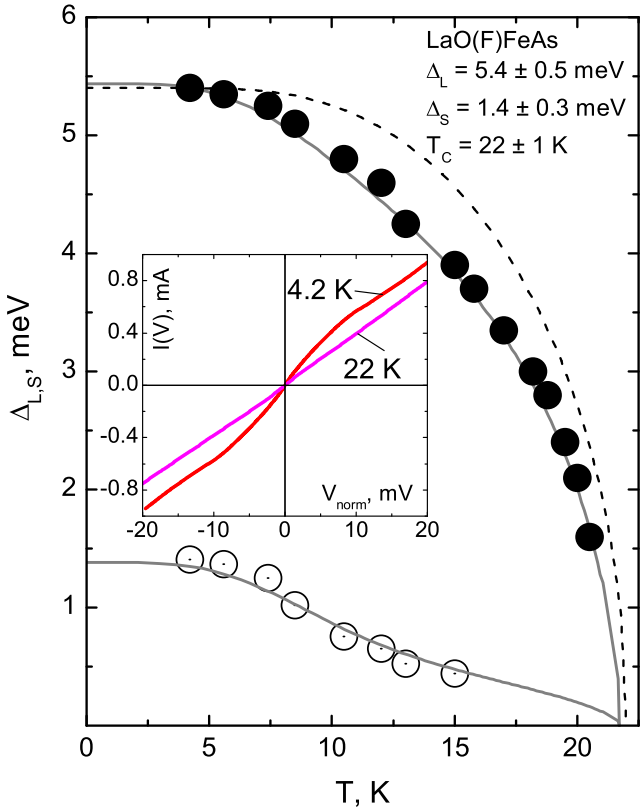}
\caption{Experimental temperature dependences for the large gap (solid circles) and for the small gap (open circles) for LaO$_{1-x}$F$_x$FeAs (see Fig.1). Solid lines show theoretical fit corresponding to the two-gap model by Moskalenko and Suhl. The inset shows current-voltage characteristics for $dI(V)/dV$ presented in Fig.1, at $T=4.2$\,K, and $T_C^{local} \approx 22$\,K.}
\end{figure}

\begin{figure}
\includegraphics[width=.49\textwidth]{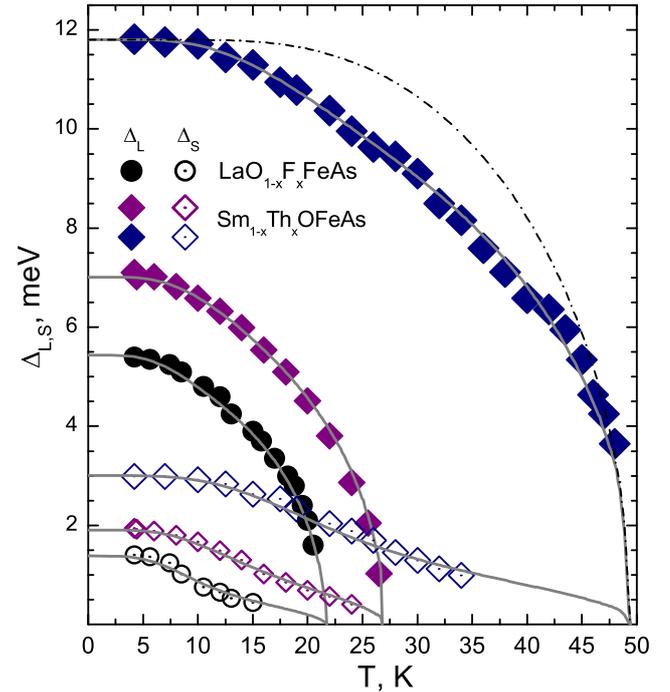}
\caption{Experimental temperature dependences for the large gap (solid symbols) and for the small gap (open symbols) for Sm$_{1-x}$Th$_x$OFeAs (rhombs), and LaO$_{1-x}$F$_x$FeAs (circles). Single-gap BCS-like curve is presented by dash-dot line for comparison. Solid lines show theoretical fitting curves determined in the framework of the two-gap model by Moskalenko and Suhl.}
\end{figure}

The temperature behaviour presented in Figs. 2,3 is typical for two-band superconductor with a moderate interband interaction \cite{MgB2fit,Nicol}. Taking into account recent ARPES studies that revealed two effective bands in 1111-oxypnictides \cite{Charnukha}, we fit the obtained $\Delta_{L,S}(T)$ by a two-band Moskalenko and Suhl gap equations \cite{Mosk,Suhl,MgB2fit} within RBCS model that allows a renormalization of $T_C^{local}$ to fit a realistic $2\Delta/k_BT_C$ BCS-ratio. The theoretical fitting curves (solid lines in Figs. 2,3) agree well with the experimental data, thus showing the simple two-band RBCS model to be appropriate for 1111 superconductors. To fit $\Delta_{L,S}(T)$ we used experimental values of $\Delta_{L,S}(4.2{\rm K})$, $T_C^{local}$, and three fitting parameters: $\alpha = \lambda_{LS}/\lambda_{SL}$, the relation between intra- and interband coupling rates $\beta = \sqrt{V_L V_S}/V_{LS}$, and the eigen critical temperature for the condensate with the small gap $T_C^S$. The only limitation for the latter parameter is evident: $2\Delta_S/k_BT_C^S > 3.52$. Red dashed lines correspond to eigen BCS-like dependences estimated by us for $\Delta_{\sigma}(T)$ and $\Delta_{\pi}(T)$   in a hypothetical case of zero interband coupling. Easy to note the condensate with the large gap looses about 14\% of eigen $T_C$ value due to interband coupling. In comparison with $\Delta_L(T)$ dependence for 1111 compounds from Fig. 3, $\Delta_{\sigma}(T)$ deviates weaker. By contrast, the bending of $\Delta_{\pi}(T)$ function is greater than $\Delta_S(T)$ for 1111, which points to weaker interband coupling rates in Mg$_{1-x}$Al$_x$B$_2$ system, probably due to the MgB$_2$ bands orthogonality in the $k$-space.

Temperature dependences of $\sigma$- and $\pi$-gaps in Al-doped Mg$_{1-x}$Al$_x$B$_2$ with $T_C^{local} \approx 21.5$\,K are shown in Fig. 4 by solid and open triangles, respectively. The temperature influence on the corresponding dynamic conductance for this contact is presented in \cite{MgB2JETPL04}. For clarity, we plot the normalized dependence $\Delta_{\pi}(T) \times \Delta_{\sigma}(0) /\Delta_{\pi}(0)$ by crossed triangles. Obviously, the $\sigma$- and $\pi$-gap behave differently with temperature increasing, thus the corresponding Andreev peculiarities in dI(V)/dV-spectra relate to distinct superconducting condensates.

\begin{figure}
\includegraphics[width=.49\textwidth]{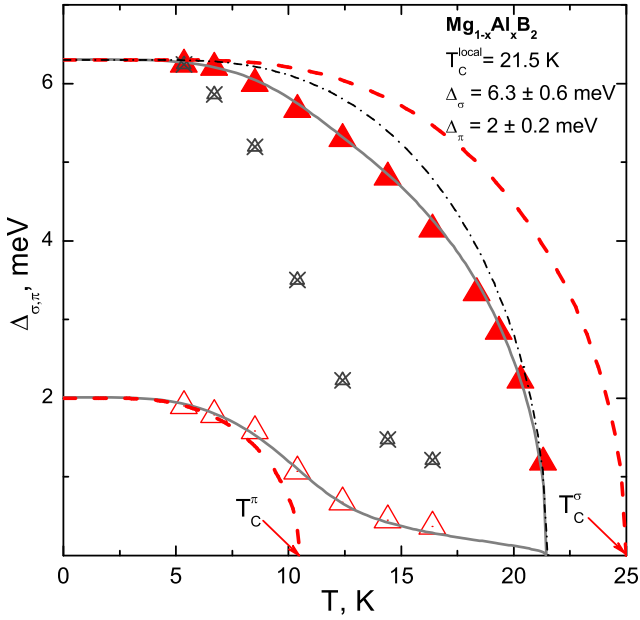}
\caption{Experimental temperature dependences for the $\sigma$-gap (solid triangles) and for the $\pi$-gap (open triangles) for Mg$_{1-x}$Al$_x$B$_2$ with $T_C^{local} \approx 21.5$\,K. Solid lines show theoretical fitting curves corresponding to two-gap model by Moskalenko and Suhl, dash-dot line show single-gap BCS-like curve. Red dashed lines correspond to eigen BCS-like dependences estimated by us for $\Delta_{\sigma}(T)$ and $\Delta_{\pi}(T)$ in a hypothetical case of zero interband coupling. The normalized dependence $\Delta_{\pi}(T) \times \Delta_{\sigma}(0) /\Delta_{\pi}(0)$ is presented by crossed triangles for comparison. The $\Delta_{L,S}(T)$ dependences were taken from \cite{MgB2JETPL04}.}
\end{figure}

From the fitting procedure, we directly estimate some important parameters of the two-gap superconducting state in GdO$_{0.88}$FeAs, Sm$_{1-x}$Th$_x$OFeAs, LaO$_{1-x}$F$_x$FeAs, and Mg$_{1-x}$Al$_x$B$_2$, summarized in Table 1. The data on LiFeAs are presented for comparison; here and below the effective value of the large gap was taken from Ref. \cite{Li13}. The gap temperature dependences for all superconductors under study demonstrate rather weak interband coupling. For oxypnictides, the eigen BCS-ratio (in a hypothetical case of zero interband coupling $V_{LS}=0$) for the large gap remains nearly constant within the whole range of critical temperatures $T_C = 22 \textendash 50$\,K, and on the average, $2\Delta_L/k_BT_C^L \approx 4.4$. This value exceeds the weak-coupling BCS-limit 3.52, thus corresponding to a strong intraband coupling in the bands with the large gap. Due to a nonzero interband interaction, the common critical temperature decreases by $~20\%$ for 1111 superconductors in comparison with the eigen $T_C^L$ for the ``driving'' bands with the large gap: in a hypothetical case of $V_{LS} = 0$, the critical temperature for oxypnictides may be as high as 70\,K. In Mg$_{1-x}$Al$_x$B$_2$ the eigen BCS-ratio for $\Delta_{\sigma}$ is larger than that in 1111, whereas the $T_C^{local}$ is closer to $T_C^{\sigma}$. The eigen BCS-ratio for the small gap tends to the BCS-limit 3.52 in both 1111 and Mg$_{1-x}$Al$_x$B$_2$ (see Table 1). The gaps ratio $\Delta_L/\Delta_S \approx 3.7$ for iron oxypnictides corresponds to the scaling between both gaps and critical temperature within the range $T_C = 22 \textendash 50$\,K \cite{UFN,FPS11}. By contrast, in Mg$_{1-x}$Al$_x$B$_2$ the $\pi$-gap does not change with temperature till $T_C \approx 15$\,K \cite{MgB2fit,MgB2JETPL04,MgB2SSC04} (which roughly corresponds to the eigen $T_C^{\pi}$ for $\pi$-bands \cite{MgB2fit,Pickett}), thus leading to $\Delta_{\sigma}/\Delta_{\pi}$ increasing.

The dependence of the relative intraband coupling constant (for the small gap) $\lambda_{22}/\lambda_{11}$ on critical temperature for 1111, Mg$_{1-x}$Al$_x$B$_2$, and LiFeAs is shown in Fig. 5. All the data follow the single dependence: $\lambda_{22}$ monotonically grows relatively to $\lambda_{11}$ with $T_C$ decreasing. Such behaviour also agrees with that predicted by the RBCS model. At the same time, the experimental points on the effective interband coupling constants, $\sqrt{\lambda_{LS} \lambda_{SL}}/\lambda_{LL}$ (open symbols in Fig. 5) obviously form two groups: $\sqrt{\lambda_{LS} \lambda_{SL}}/\lambda_{LL} = 0.08 \textendash 0.18$ for Fe-1111 compounds, and $0.03 \textendash 0.07$ for Mg$_{1-x}$Al$_x$B$_2$ . Again, the smallness of the latter constant confirms a weaker interband coupling for Mg$_{1-x}$Al$_x$B$_2$ in comparison with that for 1111.

\begin{figure}
\includegraphics[width=.49\textwidth]{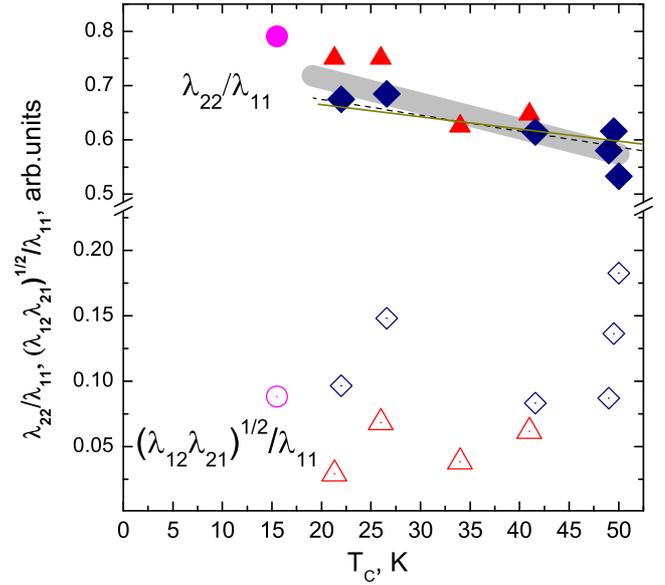}
\caption{The dependences of intraband (for the small gap) $\lambda_{22}/\lambda_{11}$ (solid symbols), and meansquare interband $\sqrt{\lambda_{12} \lambda_{21}}/\lambda_{11}$ (open symbols) relative electron-boson coupling constants on the critical temperature $T_C$ for Sm-, Gd-, and La-1111 oxypnictides (rhombs), Mg$_{1-x}$Al$_x$B$_2$ (triangles), and LiFeAs (circles) estimated in the framework of Moskalenko and Suhl equations extended for RBCS. Thin solid and dashed lines show theoretical predictions for $\lambda_{22}/\lambda_{11}$ in RBCS model. Gray area is a guideline.}
\end{figure}

In conclusion, we directly measured temperature dependences of the large and the small gap for oxypnictide superconductors GdO$_{0.88}$FeAs, Sm$_{1-x}$Th$_x$OFeAs, and LaO$_{1-x}$F$_x$FeAs, for Mg$_{1-x}$Al$_x$B$_2$, and LiFeAs. The $\Delta_{L,S}$ are well-fitted with a two-band Moskalenko and Suhl system of gap equations within the RBCS model. We estimated relative values of electron-boson coupling constants, and the eigen BCS-ratios for both bands. Our data prove a strong intraband and a weak interband coupling in the studied two-gap superconductors.

\begin{acknowledgements}
We thank Ya.G. Ponomarev, A. Charnukha for useful discussions, N.D. Zhigadlo, T. H\"{a}nke, C. Hess, B. Behr, R. Klingeler, S. Wurmehl, B. B\"{u}chner, I.V. Morozov, S.I. Krasnosvobodtsev, E.P. Khlybov, L.F. Kulikova, A.V. Sadakov, B.M. Bulychev, L.G. Sevast'yanova, K.P. Burdina, V.K. Gentchel for samples synthesis and characterization. This work was supported by RFBR Grants 13-02-01451, 14-02-90425.
\end{acknowledgements}


\onecolumn
\begin{table}[\twocolumnwidth]
\caption{Parameters of superconducting state estimated from the $\Delta_{L,S}(T)$ fitting by Moskalenko and Suhl equations. $^{\ast}$For LiFeAs, the effective large gap is taken \cite{Li13}.}
\begin{tabular}[width=1\textwidth]{|c|c|c|c|c|c|c|c|c|c|c|c|c|}
\hline
         & \multicolumn{2}{c|}{GdO$_{0.88}$FeAs} & \multicolumn{3}{c|}{Sm$_{1-x}$Th$_x$OFeAs} & LaO$_{1-x}$F$_x$FeAs & ${\rm \langle1111\rangle}$ & \multicolumn{4}{c|}{Mg$_{1-x}$Al$_x$B$_2$} & LiFeAs$^{\ast}$ \\
\hline
$T_C$, K                       & 50      & 49                    & 49.5     & 42    & 26.5                   & 22         &              & 33.7    & 21.5  & 41    &  27.2        & 15.5 \\
\hline
$\Delta_L$, meV                & 11.3    & 12                    & 11.8     & 9     & 6.4                    & 5.4        &              & 8       & 6.3   & 10     &  7.2       & 4.6 \\
\hline
$\Delta_S$, meV                & 3       & 3                     & 3        & 2.5   & 2                      & 1.4        &              & 2.2     & 2     & 2.3     & 2.1        & 1.5 \\
\hline
$\Delta_L/\Delta_S$            & 3.8    & 4                     & 3.9       & 3.6   &3.2                    & 3.9       & 3.7           & 3.6       & 3.15  & 4.3     & 3.4        & 3.1 \\
\hline
$\frac{2\Delta_L}{k_BT_C^L}$   & 4.3     & 4.8                   & 4.1      & 4.3   & 4                      & 4.6        & 4.4          & 5.2     & 5.9   & 4.8     & 5.1        & 4.8 \\
\hline
$\frac{2\Delta_S}{k_BT_C^S}$   & 3.53    & 3.8                   & 3.53     & 4     & 3.53                   & 3.53       & 3.7          & 3.9     & 4.4   & 3.53     & 3.53       & 3.7 \\
\hline
$T_C^{local}/T_C^L$            & 0.83    & 0.81                  & 0.75     & 0.85  & 0.71                   & 0.81       & 0.79         & 0.96    & 0.87  &  0.85   &  0.83       & 0.71 \\
\hline

\end{tabular}
\label{Delta}
\end{table}
\twocolumn

\end{document}